\documentclass[conference]{IEEEtran}
\IEEEoverridecommandlockouts
\usepackage{amsmath,amssymb,amsfonts,amsthm}
\usepackage{algorithmic}
\usepackage{graphicx}
\usepackage{textcomp}
\usepackage{xcolor}
\usepackage{bbold}
\usepackage{multirow}
\usepackage{balance}

\usepackage{float}
\floatstyle{plaintop}
\restylefloat{table}

\usepackage{hyperref}

\newtheorem{example}{Example}

\usepackage{siunitx}

\usepackage{svg}
\usepackage{subcaption}
\usepackage{comment}
\usepackage{listings}

\usepackage[
	backend=biber,
	style=ieee,
	sortcites=true,
	url=true,
	eprint=true,
	giveninits=false,
	minnames=1,
	maxnames=3
	]{biblatex}
\usepackage{csquotes}

\renewcommand{\baselinestretch}{.96}

\def\BibTeX{{\rm B\kern-.05em{\sc i\kern-.025em b}\kern-.08em
    T\kern-.1667em\lower.7ex\hbox{E}\kern-.125emX}}

\usepackage{tikz}
\usetikzlibrary{decorations.pathreplacing}  %
\usetikzlibrary{quantikz}

\begin{document}

\title{Joint Cutting for Hybrid Schrödinger-Feynman Simulation of Quantum Circuits
}

\author{\IEEEauthorblockN{Laura S. Herzog\IEEEauthorrefmark{2}, Lukas Burgholzer\IEEEauthorrefmark{2}\IEEEauthorrefmark{4},
Christian Ufrecht\IEEEauthorrefmark{3},
Daniel D. Scherer\IEEEauthorrefmark{3},
Robert Wille\IEEEauthorrefmark{2}\IEEEauthorrefmark{4}\IEEEauthorrefmark{5} \\}
\IEEEauthorblockA{\IEEEauthorrefmark{2}
Chair for Design Automation, Technical University of Munich, Germany}
\IEEEauthorblockA{\IEEEauthorrefmark{3}Fraunhofer Institute for Integrated Circuits IIS, Nuremberg, Germany
}
\IEEEauthorblockA{\IEEEauthorrefmark{4}Munich Quantum Software Company GmbH, Garching near Munich, Germany}
\IEEEauthorblockA{\IEEEauthorrefmark{5}Software Competence Center Hagenberg GmbH (SCCH), Hagenberg, Austria}
}

\maketitle

\begin{abstract}
    Despite the continuous advancements in size and robustness of real quantum devices, reliable large-scale quantum computers are not yet available. Hence, classical simulation of quantum algorithms remains crucial for testing new methods and estimating quantum advantage. Pushing classical simulation methods to their limit is essential, particularly due to their inherent exponential complexity. Besides the established \mbox{\emph{Schrödinger-style}} full statevector simulation, so-called \mbox{\emph{Hybrid Schrödinger-Feynman}} (HSF) approaches have shown promise to make simulations more efficient. HSF simulation employs the idea of ``cutting'' the circuit into smaller parts, reducing their execution times. This, however, comes at the cost of an exponential overhead in the number of cuts. Inspired by the domain of Quantum Circuit Cutting, we propose an HSF simulation method based on the idea of ``joint cutting'' to significantly reduce the aforementioned overhead. This means that, prior to the cutting procedure, gates are collected into ``blocks'' and all gates in a block are jointly cut instead of individually. We investigate how the proposed refinement can help decrease simulation times and highlight the remaining challenges. Experimental evaluations show that ``joint cutting'' can outperform the standard HSF simulation by up to a factor $\approx 4000\times $ and the Schrödinger-style simulation by a factor $\approx 200\times$ for suitable instances. The implementation is available at \url{https://github.com/cda-tum/mqt-qsim-joint-cutting}.
\end{abstract}


\section{Introduction}
As the availability of reliable large-scale, fault-tolerant quantum computers is still pending, classical simulation of quantum circuits remains a central tool for quantum computing research---including developing and testing quantum algorithms as well as comparing classical computers against current quantum hardware.
Such simulations are inherently challenging due to the exponential scaling of the statevectors' dimension with a growing number $n$ of qubits. The standard approach for quantum circuit simulation is the so-called \emph{Schrödinger-style} simulation which directly applies the quantum gates on the full statevector via matrix-vector multiplication. As this requires storing a vector with $2^n$ entries, storage capacities are quickly exhausted. Additionally, the runtimes increase heavily with growing $n$. Even though the exponential scaling cannot be avoided, any decrease in memory and runtime is desirable.

\emph{Hybrid Schrödinger-Feynman} (HSF)~\cite{aaronson_complexity-theoretic_2016, markov_quantum_2018, chen_64-qubit_2018} techniques tackle this by trading memory complexity for time complexity. %
This is done by partitioning large circuits into smaller subcircuits, e.g., with $\lceil n/2 \rceil $ qubits each. However, such partitioning requires ``cutting'' gates that connect the different partition which generates multiple, smaller, subcircuits to be simulated. With an increasing number of cuts, also the number of subcircuits grows exponentially. This count of smaller subcircuits, so-called ``paths'', is thus determined by the chosen quantum algorithm to be simulated---allowing faster runtimes as long as the circuit's cut induces only a reasonable number of paths.

The state-of-the-art HSF technique applies such ``cuts'' separately on each gate to be cut. We propose enhancing this procedure by, first, grouping the gates to be cut and, then, performing a ``joint cut'' on the grouped gates. Gates can be cut jointly by multiplying the respective gates and performing a Schmidt Decomposition afterwards---while this can be done automatically, one can also find analytical expressions for specific cut blocks. For certain classes of quantum circuits, the proposed method not only speeds up simulation runtimes compared to standard HSF computations but also extends the HSF simulation's applicability to instances where standard Schrödinger-style simulation would otherwise have been faster.

The remainder of this paper is structured as follows. In \autoref{sec-background} the necessary basics about quantum computing and quantum circuit simulation as well as the concepts behind HSF simulation are detailed. Afterwards, in \autoref{sec-motivation}, we introduce the idea of ``joint cutting'' and review related work. Following up on this, \autoref{sec-technical} provides important details about the realization of the proposed idea along with practical considerations. Finally, the results obtained during the evaluation of the proposed approach are summarized in \autoref{sec-usecases} and the paper is concluded in \autoref{sec-conclusion}.

\section{Background}\label{sec-background}
To keep this paper self-contained, this section reviews the basic nomenclature of quantum circuits, the core ideas of quantum circuit simulation on classical computers, and the essential concepts of Hybrid Schrödinger-Feynman simulation.
\subsection{Quantum Circuits and Schrödinger-style Simulation}

\begin{figure}[t]
    \centering
    \begin{subfigure}[c]{0.2\textwidth}
        \centering
        \includegraphics[width=\textwidth]{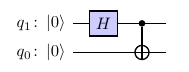} 
        \caption{}
        \label{fig:bell:subfig1}
    \end{subfigure}
    \hspace{0.05cm}
    \begin{subfigure}[c]{0.17\textwidth}
        \centering
        \includegraphics[width=\textwidth]{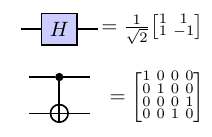} 
        \caption{}
        \label{fig:bell:subfig2}
    \end{subfigure}
    \caption{Quantum circuit (a) and the corresponding gates (b) for a Bell state preparation.}
    \label{fig:bell-prep}
\end{figure}

An $n$-qubit quantum system can be described by its \mbox{\emph{state\-vector}}, an element of a Hilbert space $\mathcal{H}_2^{\otimes n}$, where the Hilbert space of a single qubit is  $\mathcal{H}_2 = \mathrm{span}(\{\ket{0}, \ket{1}\})$. Thus, the dimension of this space scales exponentially with the number~$n$ of qubits in the considered quantum system---leading to inevitable limits regarding storing such vectors on classical machines. In Dirac notation, a state $\ket{\psi}$ residing in this space can be written as
\begin{equation}
    \ket{\psi} = \sum_{i\in\{0,1\}^n} \psi_i \ket{i},
\end{equation}
with the amplitudes $\psi_i\in\mathbb{C}$ and $i=(i_{n-1},...,i_0)\in\{0,1\}^n$. The qubits' state can be manipulated by applying quantum gates, which are described as unitary operators acting on the Hilbert space. Representing the operator in the computational basis, the output state after application of the corresponding gate can be determined by applying matrix-vector multiplication. 
Performing this procedure of consecutive matrix-vector multiplications on a classical computer is the core idea of \emph{Schrödinger-style} simulation.
We refer to~\cite{nielsen_quantum_2010} for a broader review.
\begin{example}\label{example-bell-state}
    Consider the quantum circuit shown in \autoref{fig:bell:subfig1}. The computation starts with the all-zero statevector for two qubits with $2^2=4$ entries, i.e.,
    \begin{equation}
        \ket{00} = \ket{0}\otimes \ket{0} = [1,0]^T\otimes[1,0]^T = [1,0,0,0]^T.
    \end{equation}
    Following the circuit, a Hadamard and a CNOT gate are applied, whose matrix representations are shown in \autoref{fig:bell:subfig2}. Directly performing the consecutive matrix-vector multiplication on $\ket{00}$ leads to the state
    \begin{equation}
        \bigl(\mathrm{CNOT}\cdot \bigl( (H\otimes I) \ket{00} \bigr)\bigr) = \frac{1}{\sqrt{2}} \cdot [1,0,0,1]^T,
    \end{equation}
    with $I$ being the $2\times 2$ identity matrix. The result corresponds to the well-known Bell state $\ket{\Phi}^+=\frac{1}{\sqrt{2}} (\ket{00}+\ket{11})$.
\end{example}

Classical simulation of quantum circuits, as reviewed above, can be performed by employing different data structures for representing the exponential number of amplitudes in the statevector. Directly following the treatment in \autoref{example-bell-state}, one can use plain arrays to store the statevector, matrix representations of gates, and for performing matrix-vector multiplication. Alternative structures have been investigated to cope with the exponential complexity. Examples include \emph{tensor networks}~\cite{lykov_tensor_2022, gray_hyper-optimized_2021, pednault_pareto-efficient_2020, markov_simulating_2008} which emerged from the field of condensed matter physics as well as \emph{Decision Diagrams} (DDs) originating from the \emph{Electronic Design Automation} (EDA) community~\cite{burgholzer_exploiting_2022, burgholzer_hybrid_2021, wille_tools_2022, burgholzer_tensor_2023}.

In addition to various data structures, alternative simulation schemes to standard Schrödinger-style simulation exist, as its major issue is storing and handling the exponentially large statevector. A prominent method that tackles this weakness, \emph{Hybrid Schrödinger-Feynman} (HSF) simulation, is reviewed next.

\subsection{Hybrid Schrödinger-Feynman Simulation}\label{background-HSF}
\begin{figure}[t]
	\centering
    \includegraphics[width=0.28\textwidth]{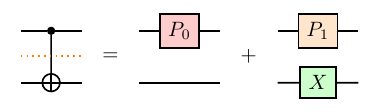}
	\caption{Bipartite representation of the CNOT gate. The orange, dotted line indicates the ``cut''.}
	\label{fig:schmidt}
\end{figure}

HSF simulation~\cite{aaronson_complexity-theoretic_2016, markov_quantum_2018, chen_64-qubit_2018} aims to reduce the inherent exponential memory complexity involved in classically simulating quantum circuits---at the expense of exponential runtime. %
The idea of HSF simulation is to horizontally partition the circuit into, for instance, two subcircuits with a similar number of qubits. This reduces the memory complexity, e.g., from $\mathcal{O}(2^n)$ to $\mathcal{O}(2^{\lceil n/2 \rceil})$, assuming a partitioning into two subcircuits with $\approx n/2$ qubits each. Since a fraction of gates usually acts across both partitions, this procedure requires ``cutting'' those connecting gates. This needs a factorized representation of the gate that ensures its constituents only act locally on their partition. The following example illustrates the idea.
\begin{example}\label{example-cnot-schmidt}
    Consider the CNOT gate as an example. Finding a bipartite representation of the CNOT gate can be easily done by factoring out properly, i.e.,
    \begin{align}
         \mathrm{CNOT} &= {\ket{0}\bra{0}} \otimes I + {\ket{1}\bra{1}} \otimes X = P_0\otimes I+P_1\otimes X.\nonumber
    \end{align}
    An illustration of this factorization is provided in \autoref{fig:schmidt}. Thus, a CNOT gate can be directly decomposed into two terms composed of local unitaries only. 
\end{example}
To find such bipartite representations more generally, one can perform a \emph{Schmidt Decomposition}~\cite{nielsen_quantum_2010} of the corresponding unitary. As will be explained in more detail in \autoref{sec-technical}, the Schmidt Decomposition allows to find representations of the form
\begin{equation}\label{equ-schmidt-decompostion-res}
    A = \sum_{m=0}^{r-1} \sigma_m\, X_m \otimes Y_m,
\end{equation}
for arbitrary operators $A:\mathcal{H}^{\otimes n}_2 \rightarrow \mathcal{H}^{\otimes n}_2$. The matrix of $X_m$ is of size $2^{n_a}\times 2^{n_a}$ while $Y_m$ is of size $2^{n_b}\times 2^{n_b}$ such that $n_a+n_b=n$ holds. The value $r$ is denoted as the \emph{Schmidt-rank} and $\sigma_m$ are the singular values.

Thus, one can find a bipartition to perform a cut for general $n$-qubit gates. Each term of the Schmidt-decomposed gate constitutes a pair of circuits to be simulated separately---each pair being a ``path'' in the overall simulation. If all gates connecting the two partitions are cut, one receives a set of smaller subcircuits, such that their memory complexity can be reduced, e.g., from $2^n$ to $2^{\mathrm{max}(n_a, n_b)}$. This reduced memory complexity then also reduces the overall runtime. The number of elements in this set, however, scales exponentially with the number of gates being cut. This leads to an overhead in time complexity, which can partly be remedied by parallel simulation of the numerous paths. The final result is reconstructed by applying the Kronecker product between the subcircuits per simulation and adding up all contributions.

The HSF technique is named for its hybrid approach, combining Schrödinger-style simulation (using matrix-vector multiplication for subcircuits) with Feynman-style simulation that explores different ``paths''. Although a full \mbox{Feynman-style} simulation, i.e., cutting \textit{all} multi-qubit gates, is computationally impractical, the hybrid method has proven quite effective.

\section{Motivation and General Idea}\label{sec-motivation}
The main strength of the HSF simulation is to reduce the memory complexity by partitioning the circuit into smaller parts. However, this comes with a cost: The more gates are cut during the partitioning, the more paths have to be simulated. Generally, the number $n_p$ of paths scales exponentially with the number of cuts. If $m$ gates are cut, one has $n_p = \prod_{i=0}^{m-1}r_i$ paths where $r_i$ is the Schmidt-rank of each cut gate. If all gate decompositions have the same rank $r$, one can simplify this to $n_p=r^m$. Thus, not every quantum circuit is suitable for being simulated with HSF approaches: Even with parallelization of the different paths, the exponentially increasing $n_p$ quickly renders the runtime prohibitive. Hence, the HSF simulation offers advantages for shallow circuits and circuits that can be partitioned into weakly connected sections---allowing for a sweet spot in the trade-off between reduced memory usage and the increased number of paths. %

\begin{figure}[t]
    \centering
    \begin{subfigure}[b]{0.26\textwidth}
        \centering
        \includegraphics[width=\textwidth]{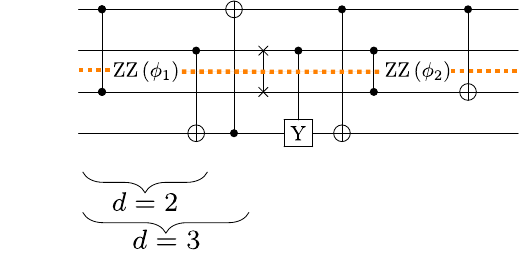}
        \caption{}
        \label{fig:pathsblock-a}
    \end{subfigure}
    \hfill
    \begin{subfigure}[b]{0.22\textwidth}
        \centering
        \includegraphics[width=\textwidth]{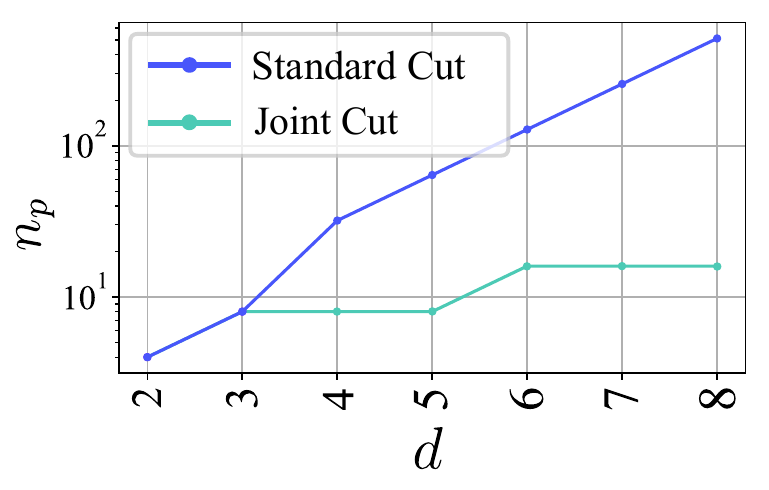}
        \caption{}
        \label{fig:pathsblock-b}
    \end{subfigure}
    \caption{(a) An exemplary circuit of which different depths $d$ are considered. For $d=2$ only the two leftmost gates are included and so on. The dotted, orange line indicates the location of the cut. (b) For different depths $d$, the number of paths $n_p$ increases more rapidly with standard cutting compared to joint cuts, which saturate. The steeper slope from $d=3$ to $d=4$ for standard cutting is due to the SWAP gate's Schmidt rank $r=4$, whereas the others have $r=2$.}
    \label{fig:pathsblock}
\end{figure}
However, in this work we show that the full potential of HSF simulation has not been reached yet. Reducing $n_p$ is possible by going beyond the naive approach of cutting each gate separately. In this work, we propose to perform ``joint cuts'' by forming suitable groups of gates in the circuit first and cutting them jointly with a Schmidt Decomposition. After illustrating how this procedure can be exploited in principle, the connections of our work and the related field of Quantum Circuit Cutting (QCC)~\cite{hofmann_how_2009,peng_simulating_2020,mitarai_overhead_2021, piveteau_circuit_2024,ufrecht_cutting_2023,ufrecht_optimal_2024,schmitt_cutting_2024} will be briefly reviewed.
\subsection{Reducing the Number of Paths with ``Joint Cutting''}
As mentioned above, individually applying cuts to each gate that connects multiple partitions directly results in an exponentially growing number $n_p$ of paths. While this exponential scaling cannot be completely avoided, there is some flexibility in the way gates are cut. For instance, if several gate cuts can be grouped, i.e., \emph{jointly} cut, the number $n_p$ of paths can be reduced considerably. An example illustrates the effect.
\begin{example}\label{example-deepblock}
Consider the circuit to be simulated as shown in \autoref{fig:pathsblock-a} and assume a partitioning that leads to a subcircuit with qubits $q_0$ and $q_1$ as well as one with qubits $q_2$ and $q_3$.
In a straightforward application of HSF simulation, one performs a Schmidt decomposition on every single gate on the cut, such that $n_p$ is the product of all separate Schmidt-ranks. Instead, we can combine the gates first and then cut them jointly by performing a Schmidt Decomposition of the resulting unitary. The behavior of $n_p$ in these cases is summarized in \autoref{fig:pathsblock-b}, where $n_p$ is plotted for increasing depths $d$ of the example circuit from \autoref{fig:pathsblock-a}. This clearly shows the potential of the ``joint cutting'' approach for which the number of paths saturates at $n_p=2^{2\cdot 2}=16$ while the state-of-the-art cutting scheme scales exponentially with $d$.
\end{example}
While this example was artificially created to demonstrate the potential benefits, our experimental evaluations, which are summarized later in \autoref{sec-usecases}, show the practical advantage of the proposed ``joint cutting'' scheme. 

The number $n_p$ of paths is reminiscent of a corresponding quantity within the field of \emph{Quantum Circuit Cutting} (QCC), which is briefly reviewed in the next section. %

\subsection{Related Work}
An advantageous aspect of (joint) cutting for quantum circuit simulation is that the \emph{Singular Value Decomposition}~(SVD) can be found for any matrix, and therefore the Schmidt Decomposition (\autoref{sec-tech-details-schmidt}) can always be constructed. Thus, the cuts can be performed fully automatically, even though it is also possible to find analytical decompositions as demonstrated in \autoref{sec-tech-joint-prac}. HSF simulation techniques enjoy further freedom since the decomposition does not need to correspond to quantum operations, hence do not need to be unitary, such as the projectors $P_0,\,P_1$ from \autoref{example-cnot-schmidt}. Changing the point of view regarding the treatment of algorithms on real quantum devices, cutting can be performed as well, but in a more constrained manner.

QCC aims to partition quantum circuits such that they can be run on smaller quantum computers. Hence, QCC and HSF approaches stem from the same core idea: Circumventing the effects of limited memory---be it RAM on a classical computer or the number of qubits in a quantum computer. Additionally, QCC can also help to reduce the impact of noise on current NISQ devices~\cite{bechtold_investigating_2023}.
Though both fields are related, decomposing gates in a QCC framework differs notably from the Schmidt decomposition applied in HSF simulations. In contrast to the aspects mentioned above, finding a decomposition for individual or combined gates cannot be done automatically and necessitates detailed, manual examination on a \mbox{case-by-case} basis.

More formally, in QCC the quantum channel of the gate is decomposed, in contrast to HSF simulation, where we decompose the matrix representing the gate. More specifically, in QCC, the decomposition takes the form of a \emph{Quasi-Probability Decomposition} (QPD) $\mathcal{W}=\sum_i a_i \mathcal{F}_i$. Each of the $\mathcal{F}_i$ is a bipartite channel, where operations are applied independently to each partition without connecting them, similar to the illustration in~\autoref{fig:schmidt}. The QPD is realized on a quantum computer by sampling the operation $\mathcal{F}_i$ with probability proportional to~$|a_i|$ and recombining the results by classical postprocessing. This procedure comes with a \emph{sampling overhead} which scales exponentially with the number of cuts. %
Note that the number of $\mathcal{F}_i$'s is reminiscent of the number $n_p$ of paths in the HSF simulation and also scales exponentially with the number of cuts in the case of QCC. An endeavor in QCC is to reduce both the number of  $\mathcal{F}_i$'s as well as the aforementioned sampling overhead. Even though the exponential scaling cannot be avoided, it can be remedied to some extent by introducing ``joint cutting''~\cite{harada_doubly_2023, ufrecht_optimal_2024, lowe_fast_2023, bechtold_joint_2024} for QCC. Similar to the proposed ``joint cutting'' for HSF simulation, in QCC, multiple gates are combined and, while exploiting the resulting structure, more beneficial decompositions can be found.

\section{Technical Details}\label{sec-technical}
In the following, we will elaborate on the ideas proposed above and illustrated in \autoref{example-deepblock}, i.e., how the Schmidt Decomposition is performed on general (blocks of) gates and why we can reach a saturation in $n_p$. Furthermore, we point out that the overhead from the Schmidt Decomposition can become significant and how ``joint cutting'' can be used in practice.
\subsection{General Schmidt Decomposition}\label{sec-tech-details-schmidt}
\begin{figure}[t]
    \centering
    \includegraphics[width=0.37\textwidth]{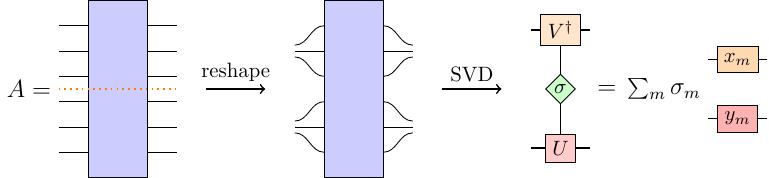}
    \caption{Illustration of reshaping the matrix of $A$ and performing an SVD.}
    \label{fig:TN_SVD}
\end{figure}
The core idea of the Schmidt Decomposition is to perform a \emph{Singular Value Decomposition} (SVD) and to rotate the basis as illustrated in \autoref{fig:TN_SVD}. As indicated in the figure, one can reshape the qubit wires (interpreted as legs of the corresponding tensor) according to the cut (orange dotted line). Afterwards, the SVD is performed which constitutes a bipartite representation. Mathematically speaking, each quantum gate is an operator acting on the qubits' Hilbert space. Given an operator $A:\, \mathcal{H}_2^{\otimes n}\rightarrow \mathcal{H}_2^{\otimes n}$ with $n$ qubits, it can be written as
\begin{equation}
    A = \sum_{i,j\in\{0,1\}^n } A_{i,j} \ket{i}\bra{j},
\end{equation}
where $i=(i_{n-1},...,i_0)$ and $j=(j_{n-1},...,j_0)$. The coefficients $A_{i,j}$ can be regarded as a $2^{2n}$-dimensional object. 
We want to decompose the unitary into partition $a$ and $b$ with the indices $i^a = (i_{n-1},...,i_{l+1}),\,i^b=(i_l,...i_0)$ and equivalently for $j^a$ and $j^b$. Thus, the cut is performed between qubit $l$ and $l+1$. This requires reshaping and matricization, i.e.,
\begin{equation}
    A = \sum_{\mathbf{i}_{\leq l}, \mathbf{i}_{>l}} \Tilde{A}_{\mathbf{i}_{\leq l}, \mathbf{i}_{>l}} \ket{i^a}\bra{j^a} \otimes \ket{i^b}\bra{j^b},
\end{equation}
with $\mathbf{i}_{\leq l}=(i^b, j^b)$ and $\mathbf{i}_{>l}=(i^a, j^a)$ as well as $ \Tilde{A}_{\mathbf{i}_{\leq l}, \mathbf{i}_{>l}}$ is a relabeled $A_{i,j}$. Now, the SVD can be applied such that
\begin{align}\label{eq-svd}
\Tilde{A}_{\mathbf{i}_{\leq l},\mathbf{i}_{>l}} = \sum_{m=0}^{r-1} U_{\mathbf{i}_{\leq l},m} \sigma_m V^*_{m,\mathbf{i}_{> l}},
\end{align}
with the \emph{Schmidt-rank} $r$. The $U$ and $V^*$ can be absorbed in the basis as
\begin{align}
    X_m &= \sum_{\mathbf{i}_{>l}} V^*_{m,\mathbf{i}_{>l}} \ket{i^a}\bra{j^a}\\
    Y_m &= \sum_{\mathbf{i}_{\leq l}} U_{\mathbf{i}_{\leq l},m} \ket{i^b}\bra{j^b},
\end{align}
leading to a bipartite representation of the original matrix of $A$ shown in \autoref{equ-schmidt-decompostion-res}. 
\subsection{Theoretical Guarantee for Lower $n_p$}
In general, it is known that the Schmidt-rank of a matrix is limited by its dimension. This means that the matrix of an operator $A : \mathcal{H}_2^{\otimes n} \rightarrow  \mathcal{H}_2^{\otimes n}$ of a block with $n$ qubits which is split into partitions with $n_a,\,n_b$ qubits, respectively, can have a Schmidt-rank at most $\Tilde{r}=\mathrm{min}(2^{2n_a}, 2^{2n_b})$~\cite{2003Schmidt-minimalterms}. This behavior is the explanation for the saturation of $n_p$ for the ``joint cutting'' in \autoref{example-deepblock}: While this bound applies individually to each cut gate when applied consecutively, it does not hold collectively for all gates. In contrast, for joint gate cutting, the limit applies to the entire block---leading to the aforementioned saturation of $2^{2\cdot 2}$. Therefore, ``joint cutting'' is guaranteed to reduce $n_p$ for deep enough blocks with limited dimensions.\\
To illustrate this in mathematical terms, consider a sequence of operators $\mathcal{N} = \{ A_k\}_{k=0}^{K-1}$ where $A_k : \mathcal{H}_2^{\otimes n} \rightarrow  \mathcal{H}_2^{\otimes n}$. They are cut into partitions with $n_a,\,n_b$ qubits respectively. The ``joint cutting'' is guaranteed to reduce $n_p$ as soon as
\begin{equation}
    n_p^t=\prod_{k=0}^{K-1} r_k > \Tilde{r} \geq n_p^{j}
\end{equation}
holds, where $r_k$ denotes the Schmidt-rank of the matrix of $A_k$. The left-hand side comes from multiplying up all \mbox{Schmidt-ranks} within $\mathcal{N}$  for the standard cutting, leading to $n_p^t$ paths. On the other hand, the maximum number of ``joint cutting'' paths, $n_p^j$, is upper bounded by $\Tilde{r}$, i.e., the smaller dimension of the reshaped matrix on which the SVD is performed (\autoref{eq-svd}).

One should note, however, that in  \autoref{example-deepblock}, already before the saturation of $n_p$, the ``joint cutting'' reduces $n_p$ in comparison to the standard cutting. Thus, the above guarantee does not need to be exhausted to make ``joint cutting'' for HSF simulations beneficial. %
\subsection{Overhead for ``Joint Cutting''}\label{sec-overhead}
Joining gates into blocks for ``joint cutting'' comes with a cost that has to be drawn into consideration regarding the usefulness of the proposed ``joint cutting'' technique. In general, merging a block of gates into one requires consecutive matrix multiplications first and afterwards a Schmidt Decomposition. The first usually has the asymptotic time complexity $\mathcal{O}(D^{s})$ with $2\leq s\leq 3$ and the SVD within the latter $\mathcal{O}(D^3)$. Here, $D=2^k$ for a $k$-qubit block, and therefore including too many qubits in a block in relation to the whole circuit can make ``joint cutting'' ineffective---unless analytical decompositions can be found (\autoref{sec-tech-joint-prac}).

Therefore \autoref{example-deepblock} may have a reduced number of paths $n_p$ but the preprocessing would dominate the total simulation runtime. Strictly speaking, the preprocessing is more costly than direct Schrödinger simulation, as the latter primarily involves \mbox{matrix-vector} multiplication ($\mathcal{O}(D^2)$). As a consequence, ``joint cutting'' can only be useful if utilized with care: Joining only a few gates in a large circuit can severely decrease the number of paths $n_p$. If small enough blocks are chosen, one receives the best of both worlds: A reduction in $n_p$ as well as negligible preprocessing overhead.

\subsection{``Joint Cutting'' in Practice}\label{sec-tech-joint-prac}
\begin{figure}[t]
    \centering
    \includegraphics[width=0.16\textwidth]{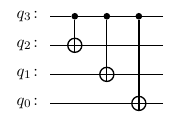}
    \caption{A ``cascade'' of CNOT gates.}
    \label{fig:cascade-cnot}
\end{figure}

In practice, it is recommended to consider quantum circuits following a beneficial structure for ``joint cutting''. One such structure is ``cascades'' of two-qubit gates. 
\begin{example}\label{ex-cascade-cnot}
    Consider \autoref{fig:cascade-cnot} in which an example cascade of CNOT gates is displayed.
    Referring to the decomposition of the CNOT gate in \autoref{example-cnot-schmidt}, one can deduce the decomposition of the cascade of three CNOT gates $C_{\mathrm{CNOT},3}$ as
    \begin{equation}
        C_{\mathrm{CNOT},3} = P_0\otimes I\otimes I\otimes I + P_1\otimes X \otimes X \otimes X.
    \end{equation}
    Thus, the Schmidt-rank remains $r=2$, such that a joint cut of this structure would be much smarter than standard cutting with $n_p^t=2^3$.
\end{example}
This structure is also beneficial for different kinds of gates such as, for instance, CZ or RZZ gates---which occur in many quantum algorithms. %

\section{Experimental Evaluation}\label{sec-usecases}
To evaluate the performance of the proposed  ``joint cutting'' method for HSF simulations, we built upon Google's qsim package~\cite{quantum_ai_team_and_collaborators_2020_4023103}, a high-performant array-based quantum simulator used for cross-entropy benchmarking in~\cite{arute_quantum_2019}. 
The ``joint cutting'' implementation is open-source and available at \url{https://github.com/cda-tum/mqt-qsim-joint-cutting}.
All evaluations were performed on a machine with a 16-core AMD Ryzen Threadripper PRO 5955WX CPU and 128 GB RAM. 
To keep the computation times reasonably low while still considering instances of relevant size, only a fraction of the total amplitudes ($10^6$) was computed for all considered methods.

\begin{figure}
    \centering
    \includegraphics[width=0.52\textwidth]{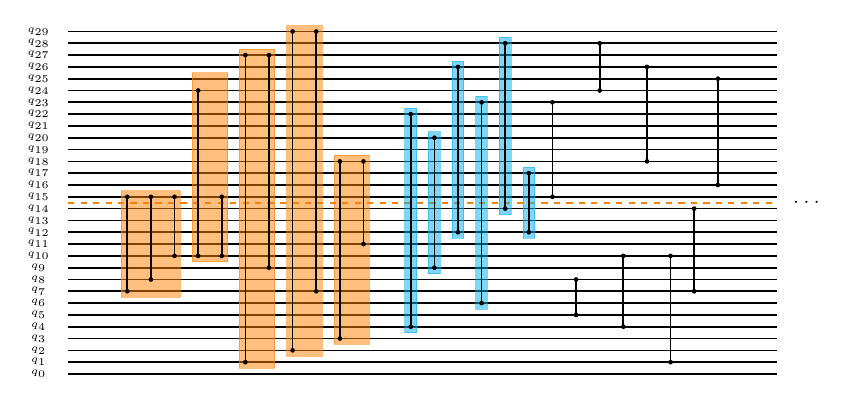}
    \caption{Examples of RZZ gates of a problem layer in QAOA from a graph with 15 qubits per partition (q30-1). The orange dotted line denotes the cut and the orange shaded gates are the utilized blocks. Blue shaded gates are cut separately. Only a few of the uncut gates are displayed without coloring.}
    \label{fig:qaoa_grouping_cascades}
\end{figure}

As a case study, we look at quantum algorithms for solving \emph{Quadratic Unconstrained Binary Optimization} (QUBO) problems, which, at the time of writing, arguably constitute the largest area of research when it comes to applications of quantum computing.
In particular, we consider the \emph{Quantum Approximate Optimization Algorithm} (QAOA) for solving the Max Cut problem, as it has been shown that any QUBO problem can be reduced to a weighted Max Cut instance~\cite{qubo-maxcut1, qubo-maxcut2, qubo-maxcut3, qubo-maxcut4}.
The circuits consist of multiple alternating so-called \emph{problem} and \emph{mixer} layers, for which one uses RZZ entangling gates as well as single-qubit RX rotation gates, respectively.
RZZ gates mutually commute, which gives a lot of freedom to reorder the gates such that grouping and, therefore, \enquote{joint cutting} can be performed---see \autoref{fig:qaoa_grouping_cascades} for an example.

To control the number of gates to cut for the HSF simulation, we considered graphs with two partitions of nodes with roughly the same size and place the cut between the partitions.
Nodes are connected with probability $p_{intra}$ within the partition and with, usually lower, probability $p_{inter}$ between the partitions. Note that an original, denser, problem graph could be shrunk to an instance that is easier to execute~\cite{herzog_improving_2024}.
Given the cut location, we use a brute force algorithm to reassemble cascades of RZZ gates which can be cut jointly. 
To keep the implementation more general, the joint cuts were performed numerically, even though analytical expressions can be found for ``cascade'' structures as shown in \autoref{ex-cascade-cnot}.

\autoref{tab:times} summarizes the results of evaluating qsim's Schrödinger Simulation, its standard, state-of-the-art HSF Simulation, and the proposed Joint HSF Simulation on various single-layer QAOA instances.
Detailed specifications of the circuits are shown in \autoref{tab:specs_qaoa_server4}.
As expected, the Schrödinger simulation's runtime mainly depends on the number of qubits---approximately doubling with every qubit added. However, for both the standard and joint HSF simulation the runtimes are heavily influenced by the number of paths. As implied in \autoref{sec-motivation}, if the sweet spot is missed, the exponentially growing number of paths can spoil the efficiency of standard HSF simulations, making it slower than Schrödinger simulation. The evaluated QAOA instances belong to this class for which HSF simulation is, in its original form, not a recommendable technique. However, the proposed ``joint cutting'' can reduce the number of paths to such an extent that it can speed up the standard HSF simulation by up to a factor $\approx 4000\times$, which, in turn, allows to outperform the Schrödinger simulation by up to a factor of $\approx 200\times$. Thus, the proposed ``joint cutting'' enables HSF techniques to be useful for instances for which the standard HSF procedure fails. This is also made possible by keeping the preprocessing costs for ``joint cutting'' low, as emphasized in \autoref{sec-overhead}. Notably, the proposed method is \textit{always} faster than the standard HSF simulation since cascades could be found in all instances.

Finally, note that ``joint cutting'' can be applied to other classes of circuits as well. 
For instance, one could simulate deeper circuits for quantum many-body dynamics as those in~\cite{richter_simulating_2024} or join CZ or iSWAP gates in shallow instances of Google's supremacy circuits~\cite{grcs_boixo}. 
We tested the latter but due to page limitations, these results are relegated to the aforementioned GitHub repository. %
\begin{table*}[t]
    \centering
    \footnotesize
    \begin{tabular}{|p{0.8cm}|p{2cm}|p{2cm}|p{1cm}|p{2cm}|p{1cm}|p{2cm}|p{2cm}|p{1.4cm}|}
    \hline
    \multicolumn{1}{|c}{} & \multicolumn{1}{|c}{Schrödinger} & \multicolumn{2}{|c}{Standard HSF} & \multicolumn{2}{|c}{Proposed Joint Cutting HSF} & \multicolumn{2}{|c|}{Performance Ratios} \\ \hline
         Circuit & Runtime (s) & Runtime (s) &  $\#$ Paths & Runtime (s)  & $\#$ Paths & S/J & T/J  \\
         \hline 
         q30-1 & $43.177\, (0.149)$ $42.915\,(0.150)$ & $118.022\, (2.041)$ $118.022\,(2.041)$ & $2^{17}$ & $0.618\, (0.021)$ $0.618\,(0.021)$ & $2^{11}$ & $\mathbf{69.857}$ & $\mathbf{190.950}$\\
         \hline
         q30-2 & $45.292\, (0.024)$ $45.029\,(0.026)$ & timed out (1 h)& $2^{26}$ & $1.480\, (0.054)$ $1.444\,(0.055)$ & $2^{12}$ & $\mathbf{30.604}$ & $\geq \mathbf{2432.544}$\\
         \hline
         q30-3 & $46.280\, (0.051)$ $46.018\,(0.051)$ & timed out (1 h) & $2^{29}$ & $49.003\, (1.648)$ $48.991\,(1.646)$ & $2^{17}$ & $0.944$ & $\geq \mathbf{73.465}$\\
         \hline 
         q31-1 & $91.956\, (0.120)$ $91.435\,(0.122)$ & $308.671\, (4.552)$ $308.671\,(4.552)$ & $2^{18}$ & $0.389\, (0.050)$ $0.379\,(0.049)$ & $2^{10}$ & $\mathbf{236.188}$ & $\mathbf{792.821}$\\
         \hline 
         q31-2 & $97.372\, (0.056)$ $96.851\,(0.056)$ & timed out (1 h) & $2^{28}$ & $13.874\, (0.122)$ $13.863\,(0.122)$ & $2^{15}$ & $\mathbf{7.018}$ & $\geq \mathbf{259.473}$ \\
         \hline 
         q31-3 & $99.860\, (0.051)$ $99.339\,(0.050)$ & timed out (1 h)& $2^{34}$ & $6.670\, (0.152)$ $6.620\,(0.146)$ & $2^{14}$ & $\mathbf{14.971}$ & $\geq\mathbf{539.726}$\\
         \hline 
         q32-1 & $201.494\, (0.039)$ $200.457\,(0.041)$ & timed out (1 h) & $2^{24}$ & $0.848\, (0.012)$ $0.822\,(0.005)$ & $2^{11}$ & $\mathbf{237.692}$ & $\geq\mathbf{4246.730}$\\
         \hline
         q32-2 & $202.769\, (0.060)$ $201.728\,(0.063)$ & timed out (1 h) & $2^{25}$ & $1.605\, (0.008)$ $1.564\,(0.007)$ & $2^{12}$ & $\mathbf{126.356}$ & $\geq\mathbf{2243.348}$\\
         \hline 
         q32-3 & $209.714\, (0.042)$ $208.676\,(0.042)$ & timed out (1 h) & $2^{30}$ & $3.219\, (0.036)$ $3.165\,(0.037)$ & $2^{13}$ & $\mathbf{65.147}$ & $\geq\mathbf{1118.324}$ \\
         \hline 
         q33-1 & $427.595\, (0.102)$ $425.519\,(0.095)$ & timed out (1 h)& $2^{24}$ & $7.409\, (0.264)$ $7.402\,(0.263)$ & $2^{14}$ & $\mathbf{57.711}$ & $\geq \mathbf{485.879}$\\
         \hline 
                           q33-2 & $435.864\, (0.183)$ $433.787\,(0.188)$ & timed out (1 h) & $2^{27}$ & $34.042\, (1.109)$ $34.031\,(1.110)$ & $2^{16}$ & $\mathbf{12.804}$ & $\geq\mathbf{105.751}$ \\
                           \hline
         q33-3 & $463.847\, (12.571)$ $457.035\,(3.128)$ & timed out (1 h)& $2^{30}$ & $16.884\, (0.100)$ $16.843\,(0.104)$ & $2^{15}$ & $\mathbf{27.473}$ & $\geq\mathbf{213.221}$\\
         \hline 

         \hline 

    \end{tabular}
    \caption{Runtimes of QAOA circuits. For the runtimes, the first line is the full time with preprocessing, and the second line is only the simulation itself. Note that the preprocessing not only contains the Schmidt Decomposition and construction of paths for the ``joint cutting'' but also gate fusion in both cases~\cite{gate-fusion, gate-fusion2}. The mean of those runtimes is displayed. In brackets, one can see the standard deviation from the mean of five runs. The column S/J shows the total \textit{S}chrödinger time divided by the full ``joint cutting'' time, and T/J the same but for the full time of standard cutting divided by the full ``joint cutting'' time. If the standard cutting is timed out, the lower bound of T/J is given. The bold entries display those in which the proposed method performs better. For all computations, the first $10^6$ amplitudes were computed.}
    \label{tab:times}
\end{table*}

\begin{table*}[t]
    \centering
    \footnotesize
    \begin{tabular}{|c|c|c|c|c|c|c|c|c|c|}
        \hline 
         Circuit & q & cut pos. & $\#$ 2-qubit gates& sizes & $p_{inter}$ & $p_{intra}$ & blocks + sep. & sep. cuts\\
        \hline
        q30-1 & 30 & 14 & 172  & [15,15] & $0.1$ & $0.8$ & 5+6 & 17 \\
        q30-2 & 30 & 14 & 181  & [15,15] & $0.15$ & $0.8$ & 6+6 & 26 \\
        q30-3 & 30 & 14 & 185  & [15,15] & 0.17 & 0.8 & 7+10 & 29\\
        \hline
        q31-1 & 31 & 14 & 186  & [15,16] & $0.1$ & $0.8$ & 6+4 & 18 \\
        q31-2 & 31 & 14 & 197  & [15,16] & $0.15$ & $0.8$ & 8+7 & 28 \\
        q31-3 & 31 & 14 & 203  & [15,16] & 0.17 & 0.8 & 9+5 & 34 \\
        \hline
        q32-1 & 32 & 15 & 206  & [16,16] & $0.1$ & $0.8$ & 6+5 & 24 \\
        q32-2 & 32 & 15 & 207  & [16,16] & $0.11$ & $0.8$ & 5+7 & 25 \\
        q32-3 & 32 & 15 & 214  & [16,16] & 0.12 & 0.8 & 7+6 & 30\\
        \hline 
        q33-1 & 33 & 15 & 219  & [16,17] & $0.1$ & $0.8$ & 6+8 & 24\\
                          q33-2 & 33 & 15 & 223  & [16,17] & $0.11$ & $0.8$ & 6+10 & 27\\
         q33-3 & 33 & 15 & 234  & [16,17] & $0.12$ & $0.8$ & 8+7 & 30\\

         \hline 
    \end{tabular}
    \caption{Specifications of the QAOA circuits where the problem graphs are generated with networkx' \lstinline{stochastic_block_model}. The column ``sizes'' indicates the number of vertices per partition of the problem graph, which sum up to $q$ and have connecting edges within the partition with probability $p_{intra}$ and between the partitions with $p_{inter}$. The ``cut pos.'' denotes the qubit label after which the cut is performed (roughly after half the qubits $q$). ``blocks + sep'' counts the number of cut blocks and the remaining separate cuts (which could not be summarized in a block). ``sep. cuts'' counts the total separate cuts. The total number of 2-qubit gates in the circuit is given in column ``$\#$ 2-qubit gates''. All QAOA instances contain a single problem layer and a single mixer layer.}
    \label{tab:specs_qaoa_server4}
\end{table*}

\balance
\section{Conclusions}\label{sec-conclusion}
In conclusion, this work presented a method for enhancing HSF simulation, enabling it to be faster than Schrödinger simulation for instances, in which the standard HSF simulation would fail. This approach is effective when circuits contain identifiable blocks (e.g., cascades), making it a valuable tool for structured circuit types. However, it is important to note that, like Quantum Circuit Cutting, HSF simulation remains best suited for specific circuit shapes, with challenges still arising for deep and dense circuits---even for the enhanced ``joint cutting''. While these limitations persist, our results highlight meaningful advancements in the scope and applicability of HSF techniques. Future work could explore further refinements in gate grouping for ``joint cutting''. For instance, in addition to regrouping the gates, adjusting the qubit order itself may help further to identify beneficial blocks that can be cut jointly.
\vspace{1cm}
\section*{Acknowledgment}
This work received funding from the European Research Council (ERC) under the European Union’s Horizon 2020 research and innovation program (grant agreement No. 101001318), was part of the Munich Quantum Valley, which the Bavarian state government supports with funds from the Hightech Agenda Bayern Plus, and has been supported by the BMWK based on a decision by the German Bundestag through project QuaST, as well as by the BMK, BMDW, and the State of Upper Austria in the frame of the COMET program (managed by the FFG).

\clearpage
\printbibliography

\end{document}